# Fragment Graphical Variational AutoEncoding for Screening Molecules with Small Data

John Armitage[a], Leszek J. Spalek[a], Malgorzata Nguyen[a], Mark Nikolka[a], Ian E. Jacobs[a], Lorena Marañón[b], Iyad Nasrallah[a], Guillaume Schweicher[a], Ivan Dimov[a], Dimitrios Simatos[a], Iain McCulloch[c], Christian B. Nielsen[b], Gareth Conduit[a] and Henning Sirringhaus[a]

In the majority of molecular optimization tasks, predictive machine learning (ML) models are limited due to the unavailability and cost of generating big experimental datasets on the specific task. To circumvent this limitation, ML models are trained on big theoretical datasets or experimental indicators of molecular suitability that are either publicly available or inexpensive to acquire. These approaches produce a set of candidate molecules which have to be ranked using limited experimental data or expert knowledge. Under the assumption that structure is related to functionality, here we use a molecular fragment-based graphical autoencoder to generate unique structural fingerprints to efficiently search through the candidate set. We demonstrate that fragment-based graphical autoencoding reduces the error in predicting physical characteristics such as the solubility and partition coefficient in the small data regime compared to other extended circular fingerprints and string based approaches. We further demonstrate that this approach is capable of providing insight into real world molecular optimization problems, such as searching for stabilization additives in organic semiconductors by accurately predicting 92% of test molecules given 69 training examples. This task is a model example of black box molecular optimization as there is minimal theoretical and experimental knowledge to accurately predict the suitability of the additives.

## Introduction

A significant attribute of organic molecules is their almost infinite chemical structural variations which exhibit a range of tunable properties[1,2]. The challenge of molecular optimization is to efficiently find the appropriate molecular structure for a particular task. In practice, while certain attributes can be simulated, this is not true for all tasks due to incomplete theory or intractable computation. In these cases, molecular optimization is driven by expensive and time-consuming empirical measurements and not by analytical predictions. A promising route to improve the searching efficiency of molecular structures is to augment computational and experimental discovery of novel materials using machine learning techniques.

Machine learning (ML) provides a route to efficiently obtain a mapping from the features of experiments to their outcomes through statistical techniques. ML algorithms have already been used to predict valid organic molecules for both pharmaceutical and organic electronics applications. These approaches focus on predicting valid molecules based on big data from either known databases or relevant theoretical calculations. For example, ML techniques are applied to big theoretical and experimental databases to predict metrics such as drug likeliness and partition coefficient, which are strong theoretical indicators to pre-screen the drugs[3–5]. In organic electronics, machine learning has been used to efficiently produce theoretical indicators for datasets which are too large to be exhaustively screened with quantum simulations[6,7]. In both of these examples, the ML is not learning from large experimental datasets but from large theoretical databases generated by accurate theoretical representations. However, very often such theoretical models do not exist and many of the existing ones are incomplete, as they generate indicators of valid molecular structures but are not able to efficiently model the complete material system. Assuming the theoretical models are valid, they can pre-screen many molecular structures; however, the remaining set of molecules, the candidate set (Figure 1), will still need to be screened, based on empirical data or expert knowledge.

[a.] *Department of Physics, Cavendish Laboratory Cambridge University, J J Thomson Avenue, Cambridge, CB3 0HE. E-mail: H.S. hs220@cam.ac.uk*
[b.] *Material Research Institute, Queen Mary, University of London, Mile End Road, London E1 4NS. E-mail: C.B.N c.b.nielsen@qmul.ac.uk*
[c.] *Solar & Photovoltaics Engineering Research Centre, Division of Physical Science and Engineering, King Abdullah University of Science and Technology (KAUST), Thuwal 23955-6900, Kingdom of Saudi Arabia.*
Electronic Supplementary Information (ESI)* available: [https://github.com/OE-FET/FraGVAE].

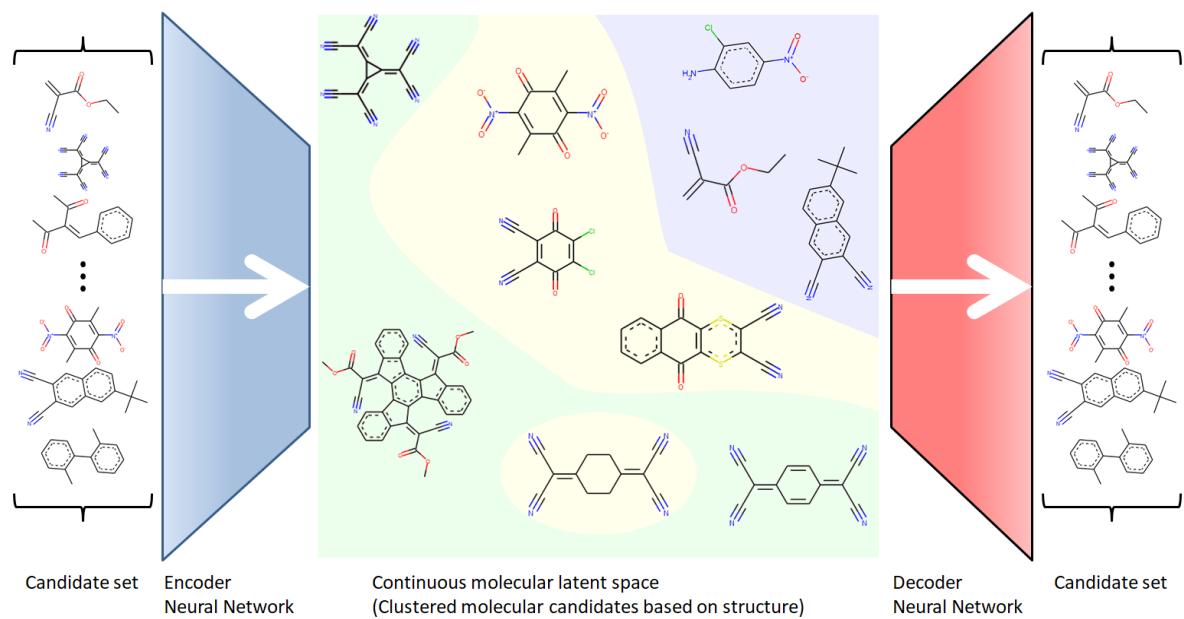

Figure 1: Pipeline of clustering molecular candidates based on structures using a molecular autoencoder. The graphical encoder reduces the dimensionality of a graph representation of a molecule to a specific point in a continuous latent space. The decoder samples the same point in the latent space to rebuild the same graph. By training the encoder and decoder to learn how to decompose and reconstruct molecules in a reduced dimensional space, the algorithm learns how to efficiently represent molecular graphs relative to other molecular graphs in the candidate set. As molecules with similar fragments are located closer in the latent space and assuming structure is related to functionality, minimal data is required to label regions of the molecular space (green and blue) in order to predict unknown regions (tan).

A clear theoretical gap in drug discovery is whether a drug has any side effects, as accurate simulations of the complete biological system are intractable. Similarly, in organic electronics, predicting the morphology of organic semiconductors purely on chemical structure is highly prone to error. Unfortunately, unlike other areas of applied machine learning without theoretical models, such as image recognition, natural language processing and finance, experimental data for molecular discovery is limited and extremely costly to acquire. As most molecular optimization problems do not have a valid and accurate theoretical model for all relevant aspects, this small data regime is the bottleneck of most molecular discovery applications and this is why molecular optimization is an extremely difficult problem. To support this effort, our objective is to provide an intuitive structural latent space based on molecular fragments, which reduces the amount of information required to find appropriate molecular structures in the candidate set. Fragments are subgraphs of molecular structures commonly used as basis functions in organic chemistry.

In machine learning, efficient encodings of data can be achieved through a process known as autoencoding, an unsupervised learning algorithm. An autoencoder consists of a neural network that learns how to copy the input to its output, however, the number of neurons representing the input at one of the layers in the pipeline is reduced, resulting in a forced dimensionality reduction, Figure 1 [8]. This technique is used in image and natural language processing to generate compressed representations of images and texts[9]. Here we train a graphical autoencoder to generate an efficient latent space representation of our candidate molecules in relation to other molecules in the set. This approach differs from traditional chemical techniques, which attempt to make a fingerprint system for all possible molecular structures instead of a specific set. Assuming a molecular structure is not randomly related to functionality, the design of a smooth structurally sorted space should also permit a smooth mapping of descriptors onto properties. This reduces the Nyquist criterion resulting in less information required to accurately model properties. Hence, a sorted space would increase the search efficiency of any black box optimization technique[10].

In summary, the primary hypothesis in this work is that graphically autoencoding candidate molecular graphs produces efficient fingerprints of candidate molecules in the small data regime. As this structure-focused approach will not be able to capture all known qualitative theoretical or experimental knowledge, this approach should be used as an unbiased quantitative structure activity relationship method to aid a collaborative decision-making process. This approach would help to augment the screening of molecular structures by providing an unbiased plausibility of subsequent molecules given the small amount of established data.

To validate our primary hypothesis that graphical autoencoded representations are appropriate for molecular fingerprints in the small data regime, we compare the predictions of our graph-based method to standard chemical and string-based molecular fingerprints in both theoretical and experimental datasets. Large theoretical datasets are used to generate robust statistics of similar small datasets under the assumption that theoretical databases are an accurate representation of practical experiments. To demonstrate graphical autoencoding in small experimental datasets, we used this technique to search for molecular additives in organic semiconductors. In organic electronics the major limiting factor for device application is that solution processed devices have poor stability. However, recent work has demonstrated that the formation of traps responsible for device degradation can be stabilized using both liquid and solid additives. As the specific mechanism is unclear, here we use ML to augment the search for new molecular additives.

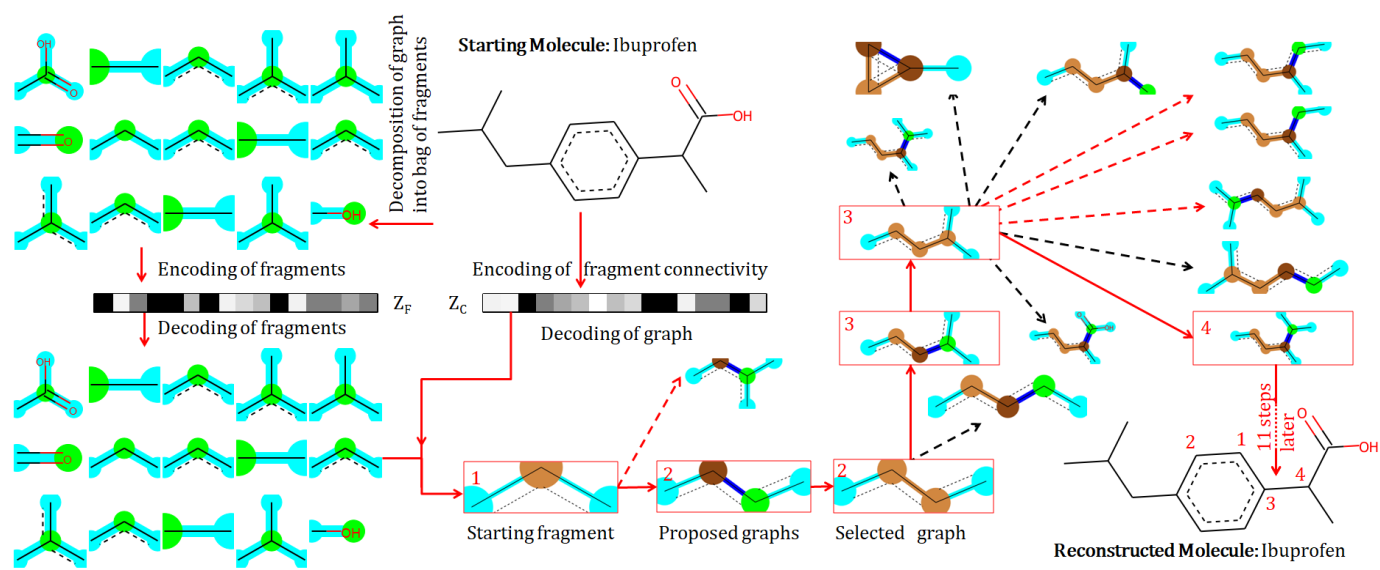

Figure 2: FraGVAE autoencoder overview: The graph is decomposed into a bag of fragments, encoded to a latent space ($Z_F$) and then decoded to reproduce the fragment bag. Secondly the connectivity of fragments is encoded to a latent space ($Z_C$) and, using the bag of fragments and $Z_C$, the molecular structure is reconstructed. Steps 1-4 demonstrate the first iterations of adding fragments/rings to a random starting fragment to reconstruct the molecular structure.

**Fragment Graphical Autoencoding**

An established approach to autoencode molecular graphs is to avoid graphs altogether and convert the graph into a one dimensional representation, such as a string based on SMILES (simplified molecular-input line-entry system) or trees [4,5]. A known problem with the string approach is that small variations in the molecular structure can result in large modifications of the string[5]. Tree structures are more robust but the encoding is still dependent on an arbitrary trace and starting node[5]. This encoding scheme results in multiple arbitrary encodings for a given molecular structure, which could result in a more complicated molecular space, undesirable for small datasets, which we avoid by encoding graphically.

Interest in the encoding of molecular graphs has exploded, resulting in a class of neural network architectures known as Message Passing Neural Networks (MPNNs), capable of generating unique encodings of molecular structure by exploiting Banach's fixed point theorem[11]. The challenge of applying these techniques to small data set applications is to ensure that the model does not overfit, as the unique encoding of larger structures requires deep MPNN. For example, to encode a fragment with a maximum degree of 3, one requires 3 graphic message passing iterations and additional layers to relate the output of the MPNN to the dependent variable. As each layer has hundreds of parameters this process is prone to over fitting[7].

To train a deep MPNN on a small dataset, here we exploit a major challenge of graphical autoencoders, direct autoencoding is intractable for reasonably sized graphs[12]. To make the problem tractable, a common approach is to perform a sequence of discrete decisions to reconstruct an undirected graph trace. In this case the next graph in the sequence is dependent only on the current state of the graph and not the history of the graph trace. This results in orders of magnitudes more training examples for the MPNN encoder for each molecular graph in the training set.

What is unique in this work is that the graph is reconstructed fragment by fragment; hence, this procedure is called Fragment Gaphical Variational AutoEncoding (FraGVAE). The smallest fragment is an extended connectivity fingerprint (ECFP) with a radius of 1, which is a node atom connected to neighboring nodes[13]. As these fragments are small, they can be directly decoded from fragment latent space ($Z_F$), unlike graphs larger than 6 nodes[14]. A property of these fragments is that each of them must be included once and only once when rebuilding the final structure to allow sampling without replacement. To reconstruct the large graph, here we randomly select a nucleating fragment with a number of dangling bonds (cyan) that can accept fragments to expand the structure. In an iterative approach, the correct neighboring fragment from the fragment bag, based on larger radius fragments indicative of the connectivity of smaller fragments in a separate latent space $Z_C$, is connected to the emerging molecule structure. The full molecular graph is encoded in the combined latent space $[Z_F, Z_C]$, Figure 1. Training a network to autoencode a molecular graph with N unique fragments which can connect to every other fragment results in N! training examples to autoencode a single molecular structure. This property helps the autoencoder to be robust to overfitting even with a small number of training examples. In contrast a similar string based approach would have 1 training example.

The decomposition of ibuprofen to circular fragments centered on atoms (green nodes) with a radius of 1 with their neighboring atoms (cyan nodes/bonds) and reconstruction process can be seen in Figure 2. This iterative process is terminated when all dangling bonds are considered, resulting in the reproduction of the original molecular structure. All dangling bonds are removed by either connecting to a terminating fragment or connecting to another dangling bond on the emerging molecule forming a ring. To check the validity of each proposed subgraph (graph index $j$) in iteration $i$, the proposed subgraph is encoded using the same encoding neural message passing network to create $Z_C$ with additional labels to atoms and bonds signifying unknown connections, which produces an encoding for $Z_{c,j}^i$. Deep learning is then used to determine the rank of the next possible subgraph $Z_{c,j}^i$ given the previous selected fragment ($Z_{c,j^*}^{i-1}$) and the connectivity of the final substructure $Z_C$.

By using circular fragments with radius 1 when decoding the connectivity, one can exploit larger circular fragments. For example, consider the combination of two fragments along an edge. Since edges are inherently one dimensional, the combination of two circular fragments with radius of 1 centered on atoms always results in a new circular fragment centered around a bond with radius 1 (Figure 3). The circular fragments centered on bonds with radius 1 must be included once and only once from a bag when rebuilding the molecular graph. As these bond fragments have set radii, undirected neural message passing networks are capable of producing unique fingerprints, similar to ECFP, each time a fragment is added to the emerging graph[15]. Interestingly, most functional groups have a set circular radius around a particular point, such as amines, sulfones, nitriles and many more. This suggests that this fragment-based scheme could be an appropriate basis set to describe a molecular functional space.

Furthermore, if each fragment centered on an atom with radius 1 is unique, knowing the set of fragments centered on bonds with radius 1 is sufficient information to reconstruct the graph. However, most fragment bags with radius 1 are not unique, therefore higher radius information is required. This would not be true if the nodes in the fragments were clusters of atoms, similar to the Jin et al. Junction Tree Autoencoder[5]. As fragment clusters tend to be unique for reasonably sized molecular graphs (30 atoms), knowing the set of unique molecular fragments centered on edges would be sufficient information to reconstruct the connectivity of the clusters. Furthermore, using clusters would prevent the transformation of emerging molecular structures converting between different topologies. Here we focus on a difficult reconstruction task of using atoms as nodes. However, it should be noted that this cluster-based fragment approach is a logical extension of this scheme and should be explored in more detail.

Unfortunately, by incorporating only a single fragment into the emerging structure, the continuous addition of fragments does not necessitate equivalent unique identifiers for larger radius fragments beyond a radius of 1 in the emerging structure due to the presence of dangling bonds compared to the encoded graph. This is exemplified by a non-circular fragment which cannot be uniquely identified as a single circular fragment around any atom or bond with any radius Figure 3. This is because the decoder does not have knowledge of the larger molecular structure beyond the dangling bond in the emerging graph. This is important as when the decoder compares the encoding of the original and emerging graph the decoder compares partial to completed circular fingerprints. This means the autoencoder learns how partial non-circular fragments are subfragments of larger circular fragments. With the assumption it is easier for the decoder to compare complete circular fragments, we bias the training data to favour the addition of fragments to bonds, which increases the maximum radius of circular fragments. However, as the molecule is generated fragment by fragment, it is trivial to determine the maximum circular radius of each fragment. To directly encode ring fragments, we incorporate an orthogonal MPNN dedicated to the communication of messages only along bonds in rings, making the network capable of generating a unique fingerprint for each ring fragment.

One outstanding issue of this approach is that it is possible to build molecularly invalid structures. For example, this method could generate a molecule with a single dangling aromatic bond and inappropriate aromatic rings. The fragment bag could also be incomplete to reconstruct a molecule. Most of these failures are avoided by hard coding rules to prevent generation and addition of certain chemically implausible fragments. To avoid an incomplete fragment bag, one could train the fragment decoder to always decode a complete set or excluding fragments which do not have a complementary fragment. A limitation of this model is that it does not contain any geometric information, hence, it is unable to distinguish stereoisomers. This could possibly be addressed by including geometric information (such as relative distance and chirality) and message passing directly to the next nearest neighbors.

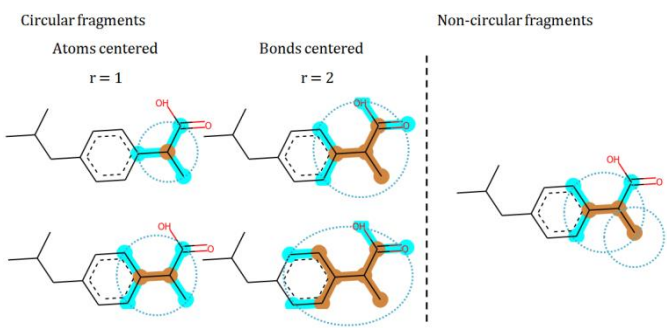

Figure 3 Examples of circular fragments centred on atoms and bonds with radius 1 and 2 and a non-circular fragment which cannot be uniquely identified from any bond/atom with a set radius. The ML algorithm learns how to describe circular fingerprints such that non-circular fragment are related appropriately.

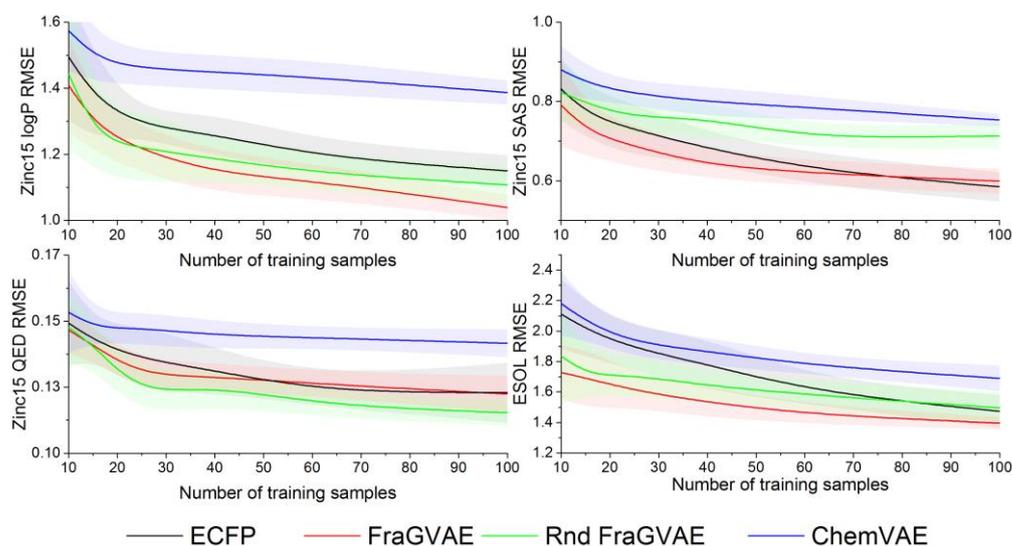

Figure 4: Predicative performance of chemical fingerprints predicting logP, SAS and QED as a function of training size dependence. The shaded area correspond to one standard deviation of the error.

Even though it is possible for the autoencoder to not decompose and reconstruct the molecule perfectly, the purpose here is to use an unsupervised learning algorithm to remove redundant information while describing the relation between fragments. In doing so this approach produces an orthogonal and complete representation of all fragments in a reduced basis set compared to standard ECFP. Due to limited computational resources we have not explored the optimization procedure for selecting model hyperparameters that trade off between completeness, orthonormality and basis set size which should be explored in future work. In this work we select intuitive dimensionality reductions and hyperparameters. For example, in our organic additives dataset we reduce the dimensionality of the ECFP with maximum radius 3 from 937 to 30 using FraGVAE which is less than the number of training examples (69).

## Results

### Predictive performance of chemical fingerprints in small calculated datasets

Here we compare the predictive performance of various chemical fingerprints trained on random small subsets (10 to 100) and tested on random larger subsets (500 to 1000) of big datasets. The molecular fingerprints methods include extended-connectivity fingerprints, ChemVAE, random FraGVAE and FraGVAE. Extended-connectivity fingerprints ECFP is a standard circular fragment based fingerprint technique used by chemists, which provides a binary identifier for each unique circle fragment of a set radius[16]. ChemVAE is the standard string based molecular autoencoder used for automatic chemical design commonly cited in the literature[4]. ChemVAE converts the simplified molecular-input line-entry system (SMILES) representations of a molecule to a one-hot encoding which is autoencoded using standard natural language processing techniques. FraGVAE with fixed random small weights were chosen, as graphical convolutions with fixed small random weights can be an appropriate fingerprint and the difference between the random and the trained FraGVAE can be attributed to the FraGVAE model learning[15].

Here we predict the theoretical octanol/water partition coefficient (logP), quantitative effective drug score (QED) and synthetic accessibility score (SAS) from 250,000 random structures from the Zinc15 database calculated by RDKIT[4,17,18]. The Zinc15 dataset was chosen as logP, QED and SAS are well established experimental indicators of suitable molecular structures with robust theoretical models[3]. In addition, most predictive performances of the autoencoded latent space models are tested on the same random Zinc15 dataset from Gomez et. al. using 10-fold cross validation [4,5,17,19,20]. Autoencoders report their predictive performance of logP, QED and SAS as they are typically used as generative models to generate new molecular structures with optimal values. In addition we also compare the experimental solubility of molecules in aqueous solutions from the ESOL datasets which are commonly used to bench mark novel MPNN in big data applications [21,22].

Prior to training the predictive models, we train FraGVAE to reconstruct molecules in the datasets, which include both the training and test data. In most ML models one always separates the training and the test set, here we train our autoencoder to reconstruct molecules in both the training and test set (candidate set). Training the autoencoder in this manor allows us to use unsupervised learning to sort the test set molecules in relation to other molecules in the training set. This approach reduces the amount of information required to find appropriate molecules in the test set.

To compare the predictive performance of chemical fingerprints trained on small datasets, we train and test a number of random forest models on random subsets of the dataset and compare the root mean squared error (RMSE), a standard practice, of different fingerprinting techniques[22]. In this small data regime we directly compare ECFP, a trained FraGVAE, FraGVAE with fixed random small weights and ChemVAE fingerprints. ChemVAE is a string based molecular autoencoder reported by Gomez et. al. [4] The maximum number of basis vectors for the FraGVAE models was selected to be on the order of a hundred dimensions which is comparable to ChemVAE and the number of training examples in the small data regime. Specific details can be seen in supplementary information*. The specific basis vectors used in each model was determined in situ by ranking the Pearson coefficient of each basis vector and selecting the top number of vectors with the largest Pearson coefficient which reduces the RMSE in threefold cross validation.

These results demonstrate that the FraGVAE fingerprints have the best predictive performance in the small data regime when predicting the logP from the Zinc15 database and the molecule solubility in the small datasets regime between 10 to 100 molecular structures compared to all other fingerprint techniques. To illustrate this point the FraGAVE model requires approximately 42 and 60 training examples to have the same error that ECFP have with 100 examples when predicting logP from Zinc15 and aqueous solubility from ESOL datasets respectively. Furthermore, the area under ECFP RMSE curve for FraGVAE is the only model that is consistently positive. It is possible the error could be further reduced by training the FraGVAE method only on the example used in the training and test set subsets instead of the complete dataset. This was not attempted as this would require a large number of FraGVAE models to develop valid statistics.

This suggests that graphical autoencoders are possibly well suited for small datasets compared to standard fingerprints, for example ECFP. This technique could be used in the large data regime as well, however, MPNN trained to directly predict a metric have been demonstrated as appropriate for large data applications when there is sufficient data to avoid over fitting[22]. FraGVAE did not reduce the RMSE error in the SAS and QED prediction in the small data regime; however, it clearly did not substantially increase the error suggesting it is a competitive fingerprint technique. For a direct comparison of the predictive performance of FraGVAE to other autoencoders in the literature in the large data regime, please see the supplementary information*.

Table 1: Area under ECFP RMSE curve between 10 and 100 training examples.

| Model | Zinc15 logP | Zinc15 SAS | Zinc15 QED | ESOL |
|---|---|---|---|---|
| ChemVAE | -18.4 | -11.8 | -1.41 | -11.9 |
| Rnd FraGVAE | 5.49 | -6.7 | 0.59 | 7.9 |
| FraGVAE | 8.26 | 1.74 | 0.06 | 17.4 |

**Screening molecular additives for organic semiconductors with neural passing network fingerprints**

To demonstrate that graphical autoencoding is a reasonable strategy in a real-world situation, we demonstrate this approach in a molecular optimization problem: searching for molecular additives for organic semiconductors. In organic electronics, the relevant material properties such as mobility, electroluminescence, quantum yield and photovoltaic efficiency are incrementally improving[23–26]. Unfortunately, the poor stability due to extrinsic environmental species, such as water and oxygen contamination, is a well-documented phenomenon that is increasingly limiting industrial applications[27]. One possible route to solve this problem is to incorporate liquid or solid-state molecular additives which improve the operational and environmental stability of conjugated polymers used in field-effect transistors and diodes[25,26,28].

The underlying mechanism of these additives is not entirely understood, but it is believed to be related to an interaction with water in voids in the polymer[25]. For solvent additives, we know that the formation of azeotropes plays a key role in removing water related traps. For solid state additives on the other hand, the mechanism is less well understood. The mechanism for the solid state additives is not clear as doping and non-doping additives improve device stability characteristics, and direct spectroscopic evidence of the additives in the film is challenging due to the small sample size and low impurity density. It is challenging to probe these voids and to determine the exact morphology of the material system, physical interactions and possible chemical byproducts to generate a clear experimental picture of the process. To find new molecular additives, there is insufficient correlational data, theoretical knowledge and experimental techniques to create an indisputable model of the material system. To find new molecular structures, one uses expert knowledge to search for new molecular additives, which can be biased. Here we augment this approach by using our FraGVAE model to provide a quantifiable unbiased opinion as to whether or not a molecular additive would improve stability in an organic electronic application.

In order to use the FraGVAE as a quantifiable unbiased opinion, 66 molecular additives were tested (Figure 5). They were chosen based on cost and variety of functional groups which are all known to show interaction with water species (ester, nitrile, phenyl, amine, nitro, quinone, sulfonic acid, ether, alcohol and halogen groups). To determine whether the additives were capable of improving device stability in organic semiconductors, they were tested in top-gate bottom-contact organic field effect transistors (OFETs), where the organic semiconductor was the amorphous polymer indacenodithiophene-co-benzothiadiazole copolymer (IDTBT).

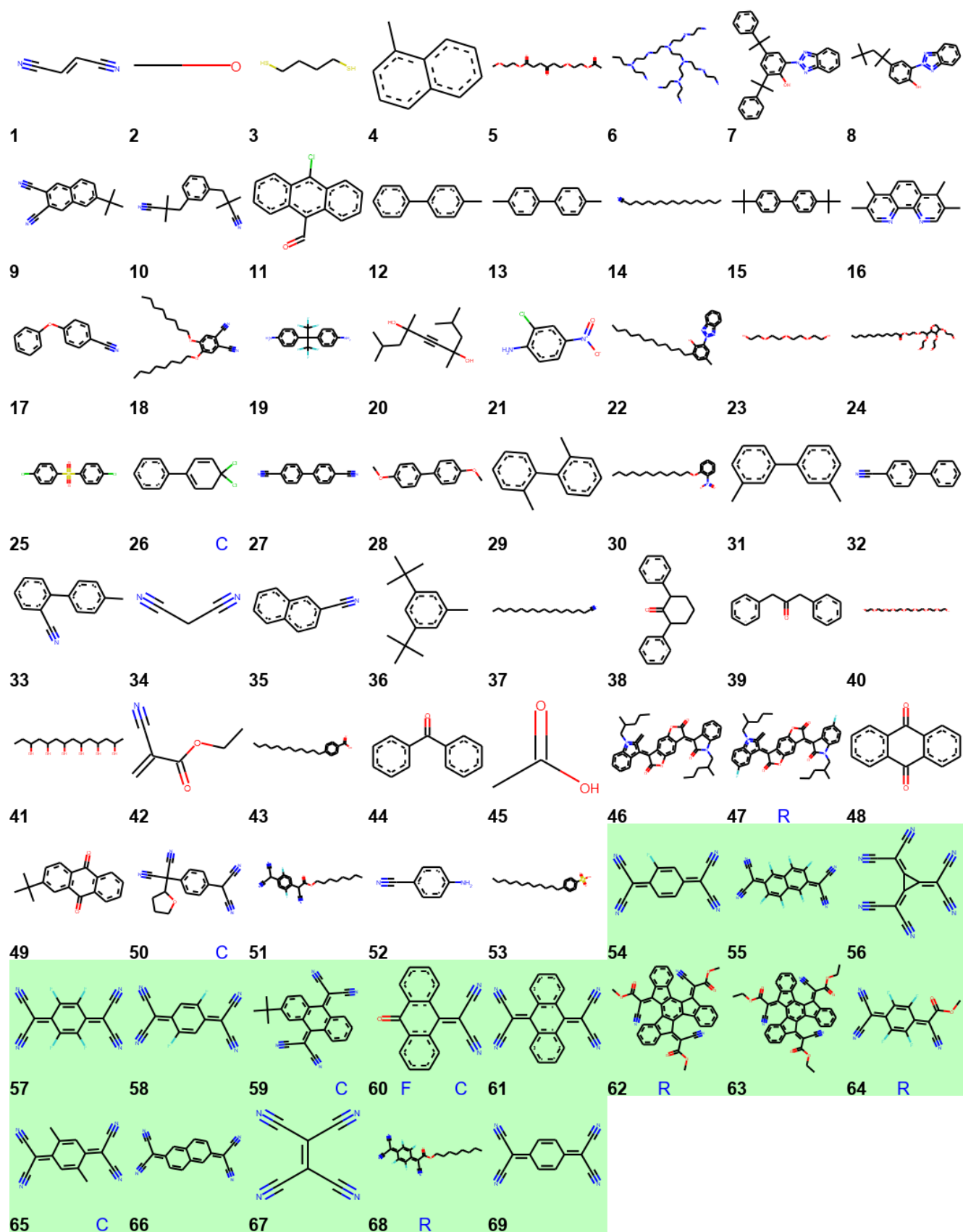

Figure 5: All additives in the cross validation set with their corresponding identification number (corresponding number appears below structure). Molecules highlighted by a green square improved device characteristics. Molecules marked with the blue characters F, R, E and C were inaccurately classified by FraGVAE, random FraGVAE, ECFP and ChemVAE respectively determined by LOOCV.

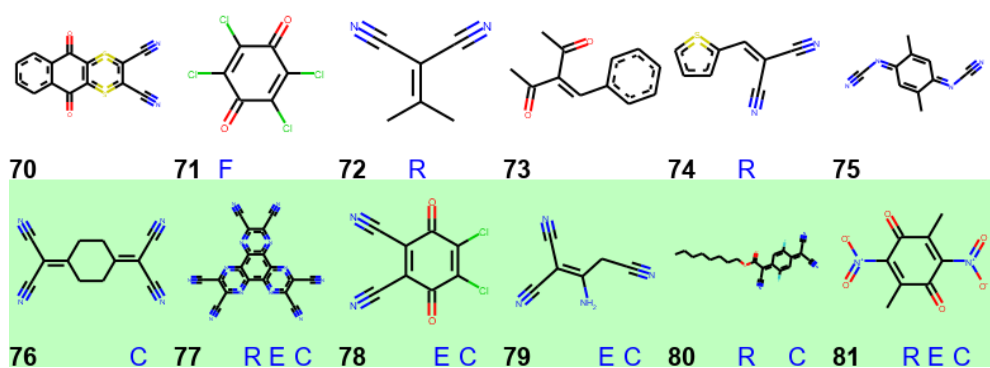

Figure 6: All molecular structures in the test set with their corresponding identification number. Molecules highlighted by a green square improved device characteristics. Molecules marked with the blue characters F, R, E and C were inaccurately classified by FraGVAE, random FraGVAE, ECFP and ChemVAE respectively.

During the fabrication, the molecular additives were incorporated into the device by blending the additive solutions into the IDT-BT solutions. Due to a large number of additives and long fabrication process, which causes variations between batches, we classified the additives in a Boolean (binary) manner to compare the results. It should be noted though, that all additives were tested against a reference and only in the case of a statistically significant improvement is an additive considered to be functional. Therefore, the additives were classified as functional additives if they were able to improve device characteristics through any process, where voltage threshold was reduced by 5V. The improvement with the addition of the additive TCNQ (red) compared to a reference sample without additives (black) is demonstrated in Figure 7, where the additive decreases the voltage threshold from -19 to -6V.

During the search for new additives, we discovered the solid-state additives undergo a chemical reaction with water, which correlates with improved device characteristics. Unfortunately, the chemical reaction is non-trivial and direct evidence of this reaction occurring in the film is challenging (more information can be seen in the supplementary section* Figures 2 to 8). As a consequence, we do not have a strong theoretical understanding of this process. Instead, we use a quantified structure-based approach as an unbiased opinion based on all empirical evidence.

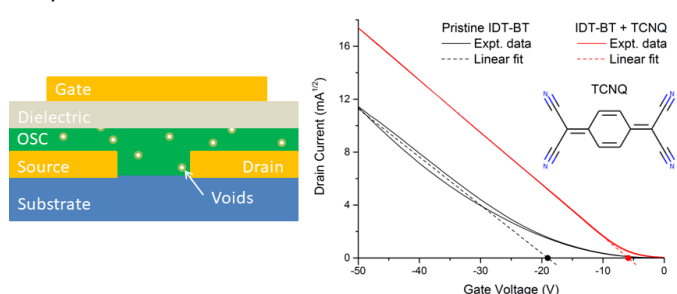

Figure 7: Top-gate bottom-contact organic transistors with voids in the organic semiconductor (OSC) film believed to be responsible for defects sites (left). Example of threshold voltage extraction from transfer characteristics of top-gate bottom-contact IDT-BT semiconductor transistor with and without TCNQ, demonstrating TCNQ is capable of improving device characteristics in OFETs (right).

Table 2: Predictive performance of different chemical fingerprinting techniques on cross validation and test set. NPV and PPV are the negative and positive predication values, i.e. the percentage of additives correctly labelled as negative and positive respectively. ROC-AUC is the area under the receiver operating characteristics curve.

|  | Cross Validation | | | Test | | |
| --- | --- | --- | --- | --- | --- | --- |
| Method | PPV | NPV | ROC-AUC | PPV | NPV | ROC-AUC |
| ChemVAE | 0.94 | 1 | 0.94 | 0 | 1 | 0.60 |
| ECFP | 1 | 1 | 0.99 | 0.33 | 1 | 0.63 |
| Rnd FragVAE | 0.94 | 0.96 | 0.94 | 0.5 | 0.67 | 0.67 |
| FragVAE | 0.94 | 1 | 0.995 | 1 | 0.83 | 0.90 |

The complete set of all molecular structures tested, their identification number and their classification of whether the molecular additive could improve devices characteristics (highlighted in green) are seen in Figure 5. Based on this small dataset, we would like to extrapolate and find new molecular structures given our small amounts of data. To extrapolate from our given data, here we train simple linear logistic regression models, which were optimized using leave-one-out cross validation. Optimization was performed via a grid search. Molecular fingerprint techniques include FraGVAE, random weight FraGVAE, ECFP and ChemVAE. To generate the FraGVAE fingerprint, we train the FraGVAE to reconstruct all molecules in both the training and test sets. For the ChemVAE model there is only a single training example for each molecular candidate, we use the ChemVAE model trained on the Zinc15 dataset along with our candidate structures.

To test the performance of the model, we selected a set of molecular additives based on chemical intuition to act as appropriate molecular additives. The molecules in the test set and their corresponding reference number and classification can be seen in Figure 7. In particular we select HAT-CN6 (#77), DDQ (#78) and Chloranil (#71), which are all well-known electron acceptors for organic electronics. The additives #72, 74, 76, 79 and 80 have the 1,1-dicyanoethene fragment observed in almost every functioning additive in the training data. In addition, we tested additives #70, 73 and 75, which have electron-accepting groups attached to quinones, which are similar structural motifs to molecular additives classified as working in the training set. We also synthesized additive # 80, which adds a soluble side chain onto F2-TCNQ (#58). The modification of the transfer characteristics with the additives can be seen in the supplementary information*. The cross validation and test metrics can be seen in Table 1 for various fingerprinting methods.

## Discussion

These results demonstrate that ChemVAE, which appears promising for Bayesian optimization of molecular structures in big data contexts, does not seem to be effective for small datasets. This is believed to be caused by the inherently discrete random jumps between near identical molecular structures due the text encoding algorithm necessity of converting an arbitrary topology of a molecular structure to a one-dimensional object. For example, the canonical SMILES representations of dimethyl-TCNQ (#65) are considerably different than the SMILES of both TCNQ (#69) and F2-TCNQ (#58). This would result in molecules with near identical structures (and assumed properties) being located in completely different locations of the latent space and more experimental data is needed to fulfil the Nyquist criterion. In big data Bayesian optimization applications, where there are appropriate theoretical models of the system, Nyquist criteria can be overcome computationally through big data.

ECFP worked extremely well on the training set, as all molecular structures in the test set can be correctly classified by identifying the presence of fragment (#72) in a ring or multiple methyl 2-cyano-3-methylcrotonate fragments (#42). This approach is over fitting so breaks down in the test set where not all additives classified as working contain the same fragment, such as additives #77, 78, 79 and 80, which have different fragments or smaller similar sub-fragments. FraGVAE was able to detect similarities between functioning additives in the training and test set even though there was no obvious fragment correlation, which is the major benefit of using graphical encoded fingerprints. By sorting the graph through MPNN, the algorithm can recognise similarities between fragments which would be ignored by standard approaches which count discrete fragments.

Experimentally the authors were surprised that additive #71 did not work as it is a known organic dopant. Interestingly FraGVAE also predicted additive #71 as an appropriate additive based on the training set. This suggests that FraGVAE intuition was reasonable to predict additive #71 as functioning even though it does not contain exactly the same functional groups that are exclusively present in positive training set examples.

## Conclusions

In this work, we address the fundamental problems in applying artificial intelligence to the majority of molecular optimization problems. The obstacle in applying ML is that there is insufficient experimental data or theoretical knowledge to build a robust statistical model to screen candidate structures. We propose an approach which uses graphical autoencoders to sort molecules based on their structures. As the graphical decoders are currently an area of interest, we propose the first fragment based decoder which reconstructs a molecular graph first through the direct decoding of small graph fragments, followed by the recombination of the fragments. Finally, we demonstrate that sorting molecular graphs with a graphical autoencoder is a valid approach to improve the predictive accuracy of quantitative structural models in the small data regime compared to standard molecular fingerprints. We have demonstrated that this method appears usually for organic electronic applications with novel materials systems which do not have an established theory and experimental practices. This approach appears promising for other fields such as drug discovery, chemistry and material science.

## Experimental

IDT-BT OFETs where Top-Gate Bottom Contact devices with a W/L of 50. All devices were prepared on Corning 1737F substrates supplied by Precision Glass & Optics. IDT-BT was supplied by I.M. and dissolved at 10g/L in DCB with the molecular additives‡. The bottom contacts were gold. IDT-BT was spun onto the substrate, baked at 90°C for 1hr. A Cytop M layer of 500nm was deposited as a dielectric. Aluminium gates were evaporated via shadow mask.

## Conflicts of interest

There are no conflicts to declare.

## Acknowledgements

J.W.A. acknowledges doctoral support from the Canadian Centennial Scholarship Fund, Christ's College Cambridge and FlexEnable.
L.J.S. is grateful to Marcin Abram for insightful discussions.
I.E.J acknowledges funding from a Royal Society Newton International Fellowship.
G.S. acknowledges postdoctoral fellowship support from the Wiener-Anspach Foundation and The Leverhulme Trust (Early Career Fellowship supported by the Isaac Newton Trust).
I.D. acknowledges NanoDTC ERC and Cambridge Philosophical society.

## Notes and references

‡ For complete table of all additives names and suppliers along with the corresponding OFET transfer characteristics please see the supplementary information*.

* Supplementary material will be released with manuscript publication.

1. G. Malliaras and R. Friend, *Phys. Today*, 2005, **58**, 53–58.
2. R. Geng, H. M. Luong, M. T. Pham, R. Das, K. S. Repa, J. Robles-garcia, T. A. Duong, H. T. Pham, T. H. Au, N. D. Lai, G. K. Larsen, M. Phan and T. D. Nguyen, , DOI:10.1039/c9mh00265k.
3. G. R. Bickerton, G. V. Paolini, J. Besnard, S. Muresan and A. L. Hopkins, *Nat. Chem.*, 2012, **4**, 90–98.
4. R. Gómez-Bombarelli, D. Sheberla, J. Aguilera-iparraguirre, T. D. Hirzel and R. P. Adams, *ACS Cent. Sci.*, 2018, 268–276.
5. W. Jin, R. Barzilay and T. Jaakkola, *arXiv*, , DOI:1802.04364v2.
6. R. Gómez-Bombarelli, J. Aguilera-Iparraguirre, T. D. Hirzel, D. Duvenaud, D. Maclaurin, M. A. Blood-Forsythe, H. S. Chae, M. Einzinger, D.-G. Ha, T. Wu, G. Markopoulos, S. Jeon, H. Kang, H. Miyazaki, M. Numata, S. Kim, W. Huang, S. I. Hong, M. Baldo, R. P. Adams and A. Aspuru-Guzik, *Nat. Mater.*, , DOI:10.1038/nmat4717.
7. J. Gilmer, S. S. Schoenholz, P. F. Riley, O. Vinyals and G. E. Dahl, *arXiv*, , DOI:10.1002/nme.2457.
8. D. E. Rumelhart and J. L. McClelland, in *Parallel Distributed Processing: Explorations in the Microstructure of Cognition: Foundations*, MIT Press, 1985.
9. P. Baldi, 2012, 37–50.
10. B. Cao, L. A. Adutwum, A. O. Oliynyk, E. J. Luber, B. C. Olsen, A. Mar and J. M. Buriak, *ACS Nano*, , DOI:10.1021/acsnano.8b04726.
11. C. Daskalakis, C. Tzamos and M. Zampetakis, .
12. L. G. May, *Constrained Graph Variational Autoencoders for Molecule Design*, .
13. R. C. Glem, A. Bender, C. H. Arnby, L. Carlsson, S. Boyer and J. Smith, *IDrugs*, 2006, **9**, 199–204.
14. G. R. U. Sing and V. A. A. Utoencoders, 2018, 1–12.
15. D. Duvenaud, D. Maclaurin, J. Aguilera-Iparraguirre, R. G´omez-Bombarelli, T. Hirzel, A. Aspuru-Guzik and R. P. Adams, *Dig. Asia-Pacific Magn. Rec. Conf. 2010, APMRC 2010*, 2015, 1–9.
16. D. Rogers and M. Hahn, *J. Chem. Inf. Model.*, 2010, **50**, 742–754.
17. T. Sterling and J. J. Irwin, *J. Chem. Inf. Model.*, 2015, **55**, 2324–2337.
18. RDKit: Open-Source Cheminformatics Software, http://www.rdkit.org.
19. M. J. Kusner, B. Paige and J. Miguel, .
20. H. Dai, Y. Tian, B. Dai, S. Skiena, L. Song and A. Financial, 2018, 1–17.
21. J. S. Delaney, 2004, 1000–1005.
22. Z. Wu, B. Ramsundar, E. N. Feinberg, J. Gomes, C. Geniesse, A. S. Pappu, K. Leswing and V. Pande, *Chem. Sci.*, 2018, **9**, 513–530.
23. J. Liu, H. Zhang, H. Dong, L. Meng, L. Jiang, L. Jiang, Y. Wang, J. Yu, Y. Sun, W. Hu and A. J. Heeger, *Nat. Commun.*, 2015, **6**, 1–8.
24. X. Gao and Z. Zhao, *Sci. China Chem.*, 2015, **58**, 947–968.
25. M. Nikolka, I. Nasrallah, B. Rose, M. K. Ravva, K. Broch, D. Harkin, J. Charmet, M. Hurhangee, A. Brown, S. Illig, P. Too, J. Jongman, I. McCulloch, J.-L. Bredas and H. Sirringhaus, *Nat. Mater.*, 2016, **1**, 356–362.
26. M. Nikolka, K. Broch, J. Armitage, D. Hanifi, P. J. Nowack, D. Venkateshvaran, A. Sadhanala, J. Saska, M. Mascal, S. H. Jung, J. K. Lee, I. McCulloch, A. Salleo and H. Sirringhaus, *Nat. Commun.*, 2019, **10**, 1–9.
27. D. M. de Leeuw, M. M. J. Simenon, a. R. Brown and R. E. F. Einerhand, *Synth. Met.*, 1997, **87**, 53–59.
28. M. Nikolka, G. Schweicher, J. Armitage, I. Nasrallah, C. Jellett, Z. Guo, M. Hurhangee, A. Sadhanala, I. McCulloch, C. B. Nielsen and H. Sirringhaus, *Adv. Mater.*, 2018, **1801874**, 1801874.